\documentclass[sigconf]{acmart}
\AtBeginDocument{%
  }

\usepackage{tabularx}
\usepackage{cleveref}
\usepackage{enumitem} % added by CMY
\begin{document}

%\title{Not all fault-errors are equal: a systematic fault analysis of an open-source RISC-V Vector Compute Cluster for Edge AI}
\title{Not All Faults Are Equal: Transient-Fault Sensitivity Characterization of an Open-Source RISC-V Vector Cluster}

\author{Maoyuan Cai}
\affiliation{%
  \institution{University of Bologna}
  \city{Bologna}
  \country{Italy}
}
\email{maoyuan.cai@unibo.it}

\author{Amirhossein Kiamarzi}
\affiliation{%
  \institution{University of Bologna}
  \city{Bologna}
  \country{Italy}}
\email{amirhossein.kiamarz2@unibo.it}

\author{Davide Rossi}
\affiliation{%
  \institution{University of Bologna}
  \city{Bologna}
  \country{Italy}}
\email{davide.rossi@unibo.it}

\author{Angelo Garofalo}
\affiliation{%
  \institution{University of Bologna}
  \city{Bologna}
  \country{Italy}}
\email{angelo.garofalo@unibo.it}

% \author{Maoyuan Cai}
% \email{maoyuan.cai@unibo.it}

% \author{Amirhossein Kiamarzi}
% \email{amirhossein.kiamarz2@unibo.it}

% \author{Davide Rossi}
% \email{davide.rossi@unibo.it}

% \author{Angelo Garofalo}
% \email{angelo.garofalo@unibo.it}

% \affiliation{
%   \institution{University of Bologna}
%   \city{Bologna}
%   \country{Italy}
% }


\begin{abstract}
  % We present a transient-fault sensitivity study of the open-source RISC-V vector cluster Spatz under SET and SEU fault models. Across 100,000 fault injections on MatMul and Widening MatMul, data-corrupting faults dominate among manifesting outcomes, making them especially critical because they are hard to detect and can severely degrade output accuracy. SET sensitivity concentrates in the vector execution subsystem, whereas SEU sensitivity is dominated by the TCDM. We further quantify SDC incidence and numerical severity across FP32, FP16, BP16, and FP8, observing at least an order-of-magnitude fewer SDC events in FP8. Exponent-targeted corruptions cause substantially larger deviations than mantissa- or sign-targeted ones, motivating selective hardening of the most critical datapaths.
  We present a transient-fault sensitivity study of the open-source RISC-V vector cluster Spatz under SET and SEU fault models. Across 100,000 fault injections on six MatMul and Widening MatMul configurations, faulty data corruption (FD) is the dominant manifesting outcome for all evaluated workloads, accounting for at least 86\% of manifesting errors in the SET campaigns and at least 91\% in the SEU campaigns. At the module level, SET sensitivity is concentrated in the vector execution path, while TCDM is the major contributor to FD manifestations. We further quantify SDC severity across FP32, FP16, BP16, and FP8 by analyzing both the average number of corrupted outputs and their RMSE. FP8 shows the lowest output impact overall, while FP16 Widening MatMul reduces both corruption spread and RMSE compared with FP16 MatMul. By contrast, the effect of widening on FP8 is limited in our experiments. Finally, exponent-targeted corruptions induce the most severe SDC events, with the largest deviations observed in FP32 and BP16, motivating selective protection of the highest-impact datapaths and fault cases.
\end{abstract}




%   \includegraphics[width=\textwidth]{Spatz cluster_drawio}
%   \caption{Spatz Cluster Overview.}
%   \Description{Spatz Cluster Overview.}


\maketitle

% Introduction
\section{Introduction}

Modern satellites and space systems in application domains such as earth observation, graceful degradation, and communication, are increasingly expected to operate as autonomous systems, rather than passive data relays. Processing data on-board reduces latency and communication costs of raw data transmission to on-earth computers. This drives a paradigm shift in on-board computers towards heterogeneous DSP-/AI-accelerated systems.

Among existing acceleration platforms, vector processors stand out as a promising paradigm because they offer a favorable balance between programmability, performance, and implementation cost, compared to other solutions. Embedded GPUs are often power-hungry \cite{geist2023nasa, kosmidis2019gpu4s}, custom accelerators and radiation-hardened FPGAs offer reduced programmability and flexibility \cite{roffe2024edgecortix}, and general-purpose CPUs typically provide insufficient throughput.

Since these processors operate in harsh radiation environments, they must be resilient to faults, modeled as Single Event Upset (SEU) and Single Event Transient (SET), the latter becoming a prominent concern in high-frequency technology-scaled-down designs. In particular, they must prevent system crashes that could cause mission failure and mitigate computation errors that could degrade AI inference accuracy. For system stability and control-flow integrity, the area overhead of conventional protection techniques on controllers and critical sections of CPUs, including radiation-hardened technology and spatial or temporal redundancy, is generally justified. However, these costs might be prohibitive to protect acceleration hardware, usually occupying a large SoC area, and must be mitigated with alternative hardware-software co-design solutions, especially when application-level error tolerance or reduced-precision data formats can inherently mask the induced errors. %warranted for the computation errors generated by the acceleration hardware which are usually large in area remains an open question, especially when application-level error tolerance or reduced-precision data formats can inherently mask the induced errors.

Addressing this point requires design sensitivity studies that take into the account algorithms, data formats, and micro-architecture. Software-only end-to-end studies capture the impact of emulated faults on AI accuracy, but miss hardware-level propagation effects~\cite{tonetto2026enfor}. Other works combine Python-based evaluation with RTL fault injection, but they fall short modeling single-event transients, and mostly target stateless GEMM accelerators~\cite{vinck2025mitigating, hoefer2023sifi, agarwal2023towards}. Other studies based on Z01X and analytical modeling provide useful insights, but do not account for system crashes and remain limited to systolic accelerators, hence not accounting for data corruption that might cause errors in the instruction-flow~\cite{tyagi2024characterizing}.

This preliminary study aims to analyze the sensitivity of an instruction-based vector processor cluster under transient fault conditions typical of space harsh environments, to derive micro-architectural design directions towards selectively protecting its critical blocks to make it suitable for space environments at the minimal hardware cost.
In summary, the contributions are:
\begin{itemize}[leftmargin=*, topsep=0pt, itemsep=0pt, parsep=0pt, partopsep=0pt]
    \item A fault-injection campaign methodology applied to an open-source RISC-V vector cluster using Synopsys VC Z01X~\cite{VC_Z01X}, covering both SEU and SET fault models and targeting the whole Spatz cluster, while enabling per-block observability to distinguish different fault manifestations, including system crashes and data corruption.
    \item An SDC-oriented fault-injection campaign on the input operands of the vector functional units under different floating-point precisions, analyzing how data format and bit-field composition (sign, exponent, and mantissa) affect fault propagation and the numerical severity of output corruption.
    \item A module-level sensitivity analysis of the Spatz cluster across representative MatMul and Widening MatMul workloads, identifying the most vulnerable components under SET and SEU fault models and deriving corresponding reliability-design guidelines.
\end{itemize}

% Background
\section{Background}
% \subsection{Scope and Assumptions}
% Our target system is a Snitch–Spatz Core Complex (CC) integrated in a shared L1 cluster with 16 SRAM banks of 8 KiB each \cite{perotti2025spatz}. The spatz cluster includes several subsystems, such as the Snitch scalar core, instruction fetch and caching structures, the shared memory system, and the Spatz vector processor. In a space environment, transient faults can affect any of these components and may ultimately influence program execution and output correctness.
% \newline While the fault surface spans the entire cluster, this paper focuses on the vector processor. Therefore, our sensitivity characterization adopts different granularities across subsystems. We model the Snitch scalar core and the instruction cache as two separate top-level blocks, but we do not further decompose Snitch into its internal microarchitectural modules. In contrast, we decompose the Spatz vector processor into its major functional blocks and perform fine-grained, per-block sensitivity analysis within Spatz. This scoping choice enables detailed attribution of transient-fault sensitivity inside the vector engine, while still accounting for the presence and contribution of non-vector components at a coarse granularity.

\subsection{Transient Fault Model}
\label{sec:fault-model}
Single Event Effects include both Single Event Transients (SETs) and Single Event Upsets (SEUs). An SET manifests as a brief voltage pulse at the output of combinational logic when an ionizing particle deposits sufficient charge in a sensitive region of a gate. Such a transient can propagate through subsequent logic and, if captured by a sequential element, may flip the stored state and appear as a soft error, namely an SEU. Soft errors can also arise when energetic particles directly disturb storage elements such as flip-flops and latches, causing a state change without requiring prior propagation through combinational paths \cite{tyagi2024characterizing}.

\subsection{Spatz Cluster and Scope of Analysis}
The Spatz cluster~\cite{perotti2025spatz} is an open-source RISC-V vector-processing cluster built around the Spatz vector processor, which implements the RVV Zve64d ISA extension. In this work, we use a configuration with one core complex (CC), integrating the scalar core \texttt{Snitch}, the tightly coupled data memory (\texttt{TCDM}), and the main vector-processing back-end of Spatz. \texttt{Snitch} is responsible for instruction decoding and offloading vector operations, while the vector back-end comprises the controller, \texttt{VFU}, \texttt{VLSU}, \texttt{VSLDU}, and \texttt{VRF}. The \texttt{VLSU} handles vector memory accesses, the \texttt{VSLDU} supports permutation operations, and the \texttt{TCDM} provides shared low-latency data storage for the cluster components.

The \texttt{VRF} is a centralized vector register file supporting three operand reads and one result write per cycle (3R1W), implemented using two SCM banks in which each of the 32 architectural vector registers occupies one row of 32\,B~\cite{perotti2025spatz}.

In our platform, the \texttt{VFU} integrates eight FPUs based on FPnew~\cite{mach2020fpnew}, an open-source transprecision floating-point unit architecture supporting multi-precision floating-point computation with packed-SIMD execution. In this work, we perform the reliability analysis at the cluster level while focusing on the Spatz vector processor. We use \textit{MatMul} and \textit{Widening MatMul} as representative workloads because they strongly exercise the floating-point vector datapath.

% Methods
\section{Methods}
Although transient faults in any cluster component can affect execution correctness, the main objective of this work is to assess the reliability of the Spatz vector processor. Therefore, while fault injection is performed at the cluster level, the sensitivity analysis adopts different granularities across subsystems: \texttt{Snitch} and the instruction cache are treated as separate top-level blocks, and we perform a finer-grained analysis at functional-unit level for the vector co-processor system.

To obtain soft-error sensitivity data, we use RTL fault injection with Synopsys VC Z01X \cite{VC_Z01X}. It performs concurrent fault simulation by tracking divergences between a fault-free good machine (GM) and faulty machines (FMs), enabling efficient campaign-scale evaluation at RTL.
We consider the fault model introduced in~\ref{sec:fault-model}, including both SETs and SEUs~\cite{tyagi2024characterizing}. 
We execute the analysis on key AI kernels: matrix multiplication (MatMul), where all matrices feature the same precision $n$-bits, and widening multiplication (widening MatMul), where output matrix is $2n$-bits. To analyze the effects of data formats on fault-error propagation in the micro-architecture and how their different range can inherently tolerate faults, we span different FP precisions, from FP32 down to FP8. 
% Since each FPU in Spatz is 64-bit with packed-simd support~\cite{bertaccini2022minifloat}, we exploit this datapath in our analysis for all precisions. 

%We use matrix multiplication (MatMul) and widening matrix multiplication (Widening MatMul) as representative workloads, since matrix multiplication is a dominant compute kernel in SLM-relevant execution and captures the main compute pattern exercised by Spatz \cite{perotti2025spatz,purayil2025troop}. These kernels further activate the FPnew \cite{mach2020fpnew}-based floating-point datapath and its packed-data SIMD support \cite{bertaccini2022minifloat}, making them suitable for sensitivity characterization. While not exhaustive, they provide a representative and reliability-relevant use of the underlying compute substrate.

%FI Campaign Configuration --> TAble
% \input{tables/tab1_FI_campaign_cofiguration}

\begin{figure}[t]
  \centering
  \includegraphics[width=0.8\linewidth]{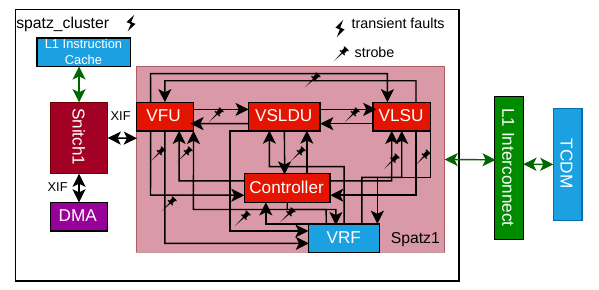}
  \caption{Strobe Positions of Spatz.}
  \label{fig:Spatz_FI}
  \Description{Strobe Positions of Spatz.}
\end{figure}

\textbf{Error detection, classification, and observability} 
By analyzing the micro-architecture of the Spatz cluster, we distinguish two error classes, corresponding to different system-level consequences: 
\begin{itemize}
\item {\texttt{Faulty System Crash (FS)}}: An error that occurs in the handshake signals between modules (e.g. between VRF and the Vector controller) might cause deadlock and induce a system crash.
\item {\texttt{Faulty Data Corruption (FD)}}: An error affecting functional data values exchanged between modules, such as operands, intermediate values, or output results, is considered a data corruption.
% \item {\texttt{Faulty Control Logic (FC)}}: An error that results in the incorrect execution of instructions, e.g. altering the program counter or generating a wrong address for memory accesses, potentially affecting the control flow of the system.
\end{itemize}

As shown in \cref{fig:Spatz_FI}, we add strobes for FS and FD on the connection ports of the inner-Spatz modules and the output port of the Spatz core. These strobes serve as the VC Z01X observation points for comparing the GM and FM at every negative clock edge during kernel execution. Detected mismatches are then used to classify the run outcome. Faulty simulations that do not trigger any of these error classes are considered \texttt{Masked}.

Of the fault classes identified above, \texttt{FD} can propagate through the computation without triggering any \texttt{FS} condition, silently generating data errors affecting the executed kernels' or programs' results. These are indicated as \texttt{Silent Data Corruption (SDC)}~\cite{ocp_sdc_ai_2025}, potentially leading to non-negligible accuracy degradation in AI workloads. Therefore, the rate and the numerical impact of such corruptions become a primary concern of the analysis.
%In addition to the above execution-time observability, we use \texttt{Silent Data Corruption (SDC)} as an \emph{end-to-end} outcome: SDC is declared only when the kernel completes without triggering an FS condition, yet the final architectural output differs from the golden result .

% \textcolor{red}{Here we should also mention the kernels we used and why we used those kernels. (For example saying that matmul, dotp, etc are kernels widely used in application scenarios where Spatz will be used).}

\begin{figure}[tb]
  \centering
  \includegraphics[width=0.8\linewidth]{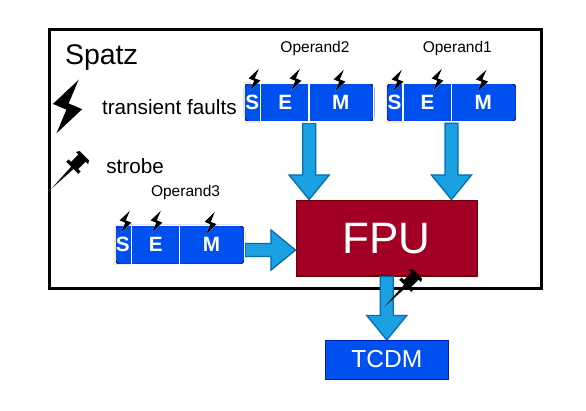}
  \caption{SDC Detection Methodology. S/E/M denote sign, exponent, and mantissa, respectively.}
  \label{fig:SDC_FI}
  \Description{SDC detection methodology.}
\end{figure}
\textbf{SDC-oriented sensitivity metric} In addition to module-level sensitivity, we characterize \texttt{SDC} from a numerical perspective. Specifically, we study how SDC depends on (i) the \emph{data precision} used by the kernel (FP32/FP16/BP16/FP8) and (ii) the \emph{floating-point value composition}, i.e., whether corruption manifests predominantly in the sign, exponent, or mantissa field. Reduced precision changes both field encoding and dynamic range, altering how faults translate into output deviations. The FI framework for SDC-oriented sensitivity evaluation is shown in ~\cref{fig:SDC_FI}.
For each precision, we interpret the operands presented to the FPUs in the VFU according to the corresponding numeric encoding and conceptually decompose each value into sign ($s$), exponent ($e$), and mantissa ($m$) fields.

To assess the sensitivity of different floating-point encoding components, we perform 1{,}000 targeted fault-injection trials per component ($s/e/m$) for each precision. In each trial, we flip exactly one bit in the \textbf{input operands to the FPU}, at a randomly selected bit-level location within the selected component and at a randomly selected clock cycle within the kernel execution window, while leaving all other bits unchanged. The sampling is uniform over all eligible bit-level locations and injection times.
During the execution of each workload, we first apply the \texttt{FS} detection described in Sec.~3.2: if a deadlock or an error on handshake signals is observed, the run is classified as \texttt{FS}. 

For runs that do not trigger \texttt{FS}, we enable a strobe in the output write-out phase of the result matrix and perform a bitwise comparison between the GM and FM outputs. If any output element differs, the run is classified as \texttt{SDC}. 
Beyond counting SDC occurrences, we quantify the numerical impact of SDC on the final outputs. Let $\mathbf{b}^{GM}$ and $\mathbf{b}^{FM}$ denote the golden and faulty output \emph{bit patterns}, where each element has width $w\in\{32,16,8\}$ bits depending on the precision. Each element is mapped to a real-valued number through a precision-specific decoding function $\mathcal{D}_{p}(\cdot)$: FP32 and FP16 follow IEEE-754 decoding, BP16 follows E8M7 decoding, and FP8 follows E5M2 decoding. This yields real-valued outputs $\mathbf{y}^{GM}=\mathcal{D}_{p}(\mathbf{b}^{GM})$ and $\mathbf{y}^{FM}=\mathcal{D}_{p}(\mathbf{b}^{FM})$, where $y^{GM}_i=\mathcal{D}_{p}(b^{GM}_i)$ and $y^{FM}_i=\mathcal{D}_{p}(b^{FM}_i)$.

For each SDC run, we define the set of corrupted output elements $\mathcal{I}=\{i \mid b^{FM}_i \neq b^{GM}_i\}$ and its cardinality $K=|\mathcal{I}|$. Here, $K$ varies across runs because a single injected fault can corrupt one or multiple output elements, depending on where and when the fault is injected and how it propagates through the kernel computation. We then quantify the numerical severity of SDC using the root mean square error (RMSE) over $\mathcal{I}$:
\begin{align}
\mathrm{RMSE} &= \sqrt{\frac{1}{K}\sum_{i\in\mathcal{I}}\left(y^{FM}_i-y^{GM}_i\right)^2}.
\end{align}

% Results
\section{Results}
% This section presents the results of our RTL fault-injection campaigns. We first analyze module-level sensitivity to show where faults most frequently manifest in Spatz. We then examine SDC sensitivity by data precision and floating-point bit-field to show how Spatz's sensitivity to SDC is shaped by these factors.
\subsection{Module-level sensitivity}
This subsection presents the module-level fault sensitivity analysis of Spatz under the considered fault models and workloads. The evaluation covers six workload configurations: FP32 MatMul, FP16 MatMul, BP16 MatMul, FP16 Widening MatMul, FP8 MatMul, and FP8 Widening MatMul. We distinguish the results by fault type, considering \textbf{SEU} injections on sequential elements (\textit{flops}) and \textbf{SET} injections on combinational logic (\textit{ports}). For each injected fault site, the run outcome is classified as FS or FD, according to the strobes defined in Sec.~3.

\begin{figure}[t]
  \centering
  \includegraphics[width=\linewidth]{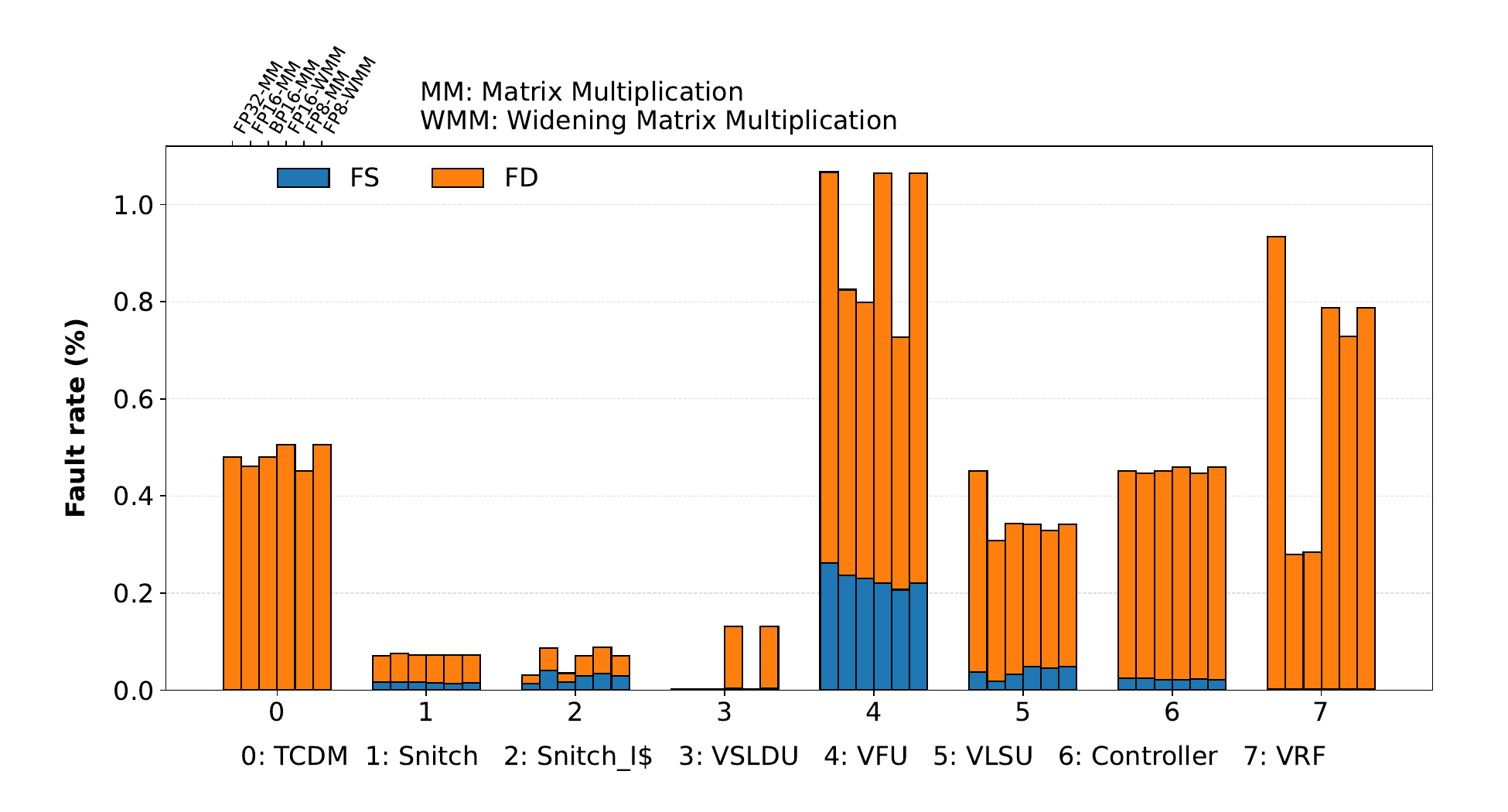}
  \caption{Module sensitivity to SETs.}
  \label{fig:SET_FI_results_general}
  \Description{General SET FI Test Results of Spatz.}
\end{figure}

\begin{figure}[t]
  \centering
  \includegraphics[width=\linewidth]{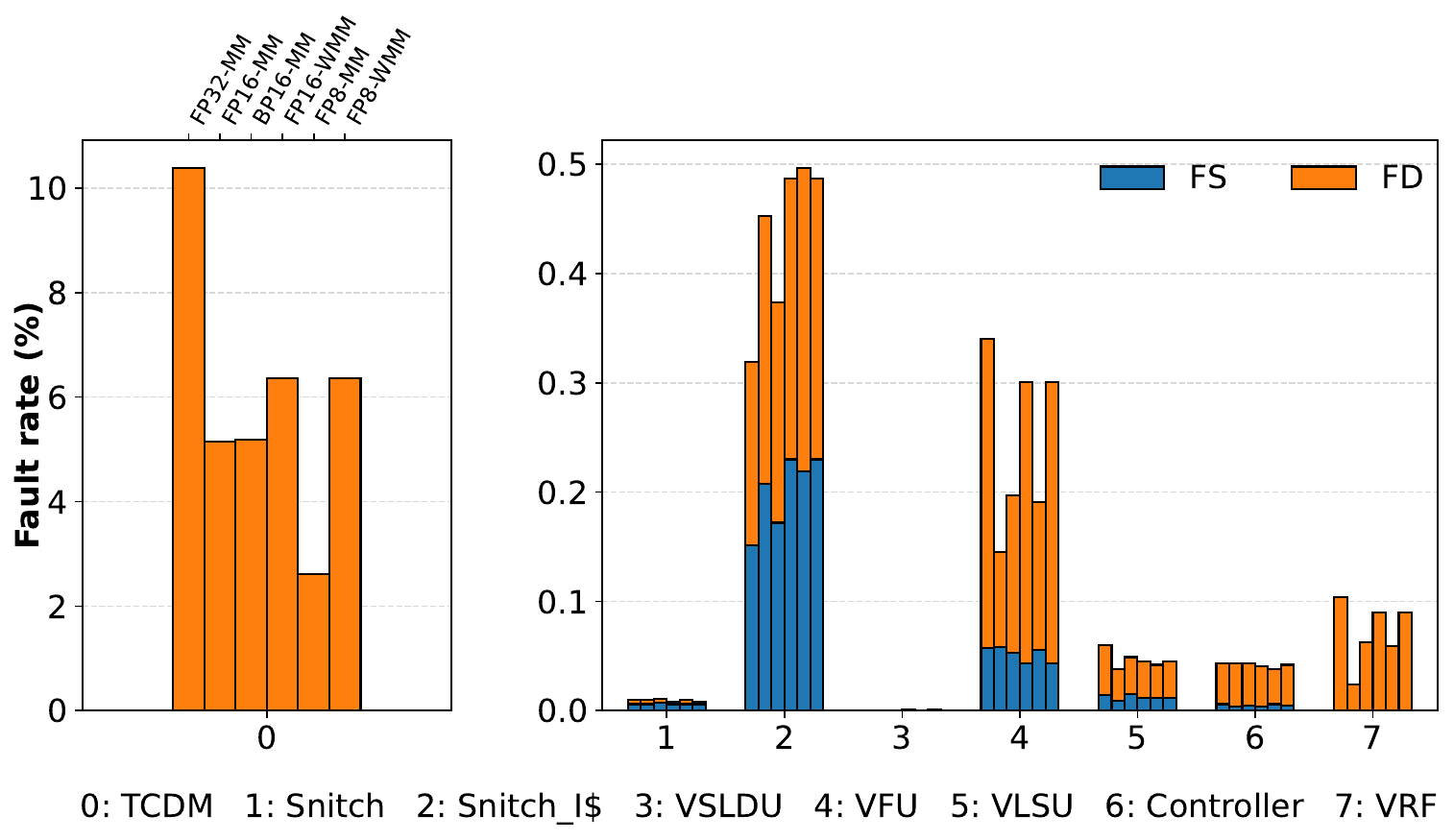}
  \caption{Module sensitivity to SEUs.}
  \label{fig:SEU_FI_results_general}
  \Description{General SEU FI Test Results of Spatz.}
\end{figure}

The results, reported in \cref{fig:SET_FI_results_general} and \cref{fig:SEU_FI_results_general}, highlight different sensitivity patterns across modules and fault manifestations.

At the module level, SET sensitivity is concentrated in the vector execution path: the \texttt{VFU}, \texttt{VLSU}, \texttt{controller}, and \texttt{VRF}-related datapaths account for at least 75\% of all manifesting SET errors, while SEUs mostly manifest on the \texttt{TCDM}, which contributes at least 75\% of all manifesting SEU errors across all evaluated workloads. Around 0.25\% of SETs and 0.05\% of SEUs manifest as \texttt{FS} on the handshake ports of the \texttt{VFU}. Since these handshake paths are implemented with lightweight control signals, protecting them through duplication or similar low-cost redundancy can significantly reduce system-crash rates with limited area overhead. A similar design consideration also applies to the handshake and control paths of the \texttt{VLSU}. Storage structures, especially \texttt{TCDM} and \texttt{VRF}, dominate \texttt{FD} manifestations, highlighting the importance of ECC/parity and periodic scrubbing in reliability-oriented design. The \texttt{Snitch} instruction cache also contributes substantially to both \texttt{FS} and \texttt{FD}, suggesting that ECC/parity and periodic scrubbing can mitigate both crash-inducing and data-corrupting faults. As for the \texttt{Snitch} scalar core, its relatively small area footprint~\cite{perotti2025spatz} makes duplication a potentially cost-effective hardening option.

Across all evaluated workloads, \texttt{FD} remains the dominant outcome under both fault models. In particular, \texttt{FD} accounts for at least 86\% of all manifesting errors in the SET campaigns and at least 91\% in the SEU campaigns. This indicates that, once a fault escapes masking, it is much more likely to corrupt data than to trigger an immediately visible failure, making silent propagation through the datapath the primary reliability concern. Therefore, the next subsection further analyzes \texttt{SDC} from a numerical perspective.

\subsection{SDC sensitivity by precision and bit-field}
\begin{figure}[t]
  \centering
  \includegraphics[width=0.9\linewidth]{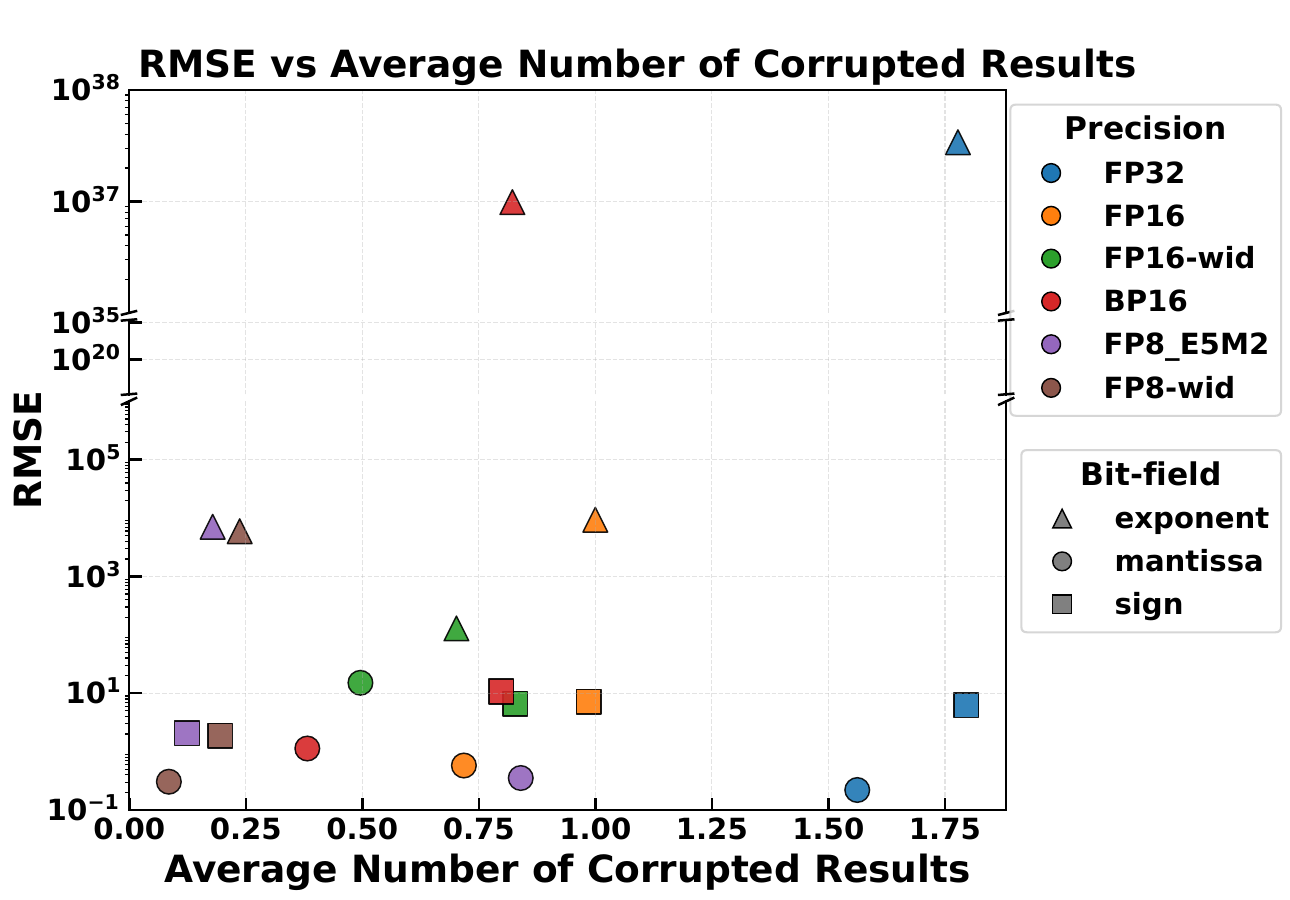}
  \caption{RMSE versus average number of corrupted outputs for manifested \texttt{SDC} runs. Note: A single injected fault can corrupt multiple output elements.}
  \label{fig:FI_results_SDC}
  \Description{Element-wise SDC severity by precision and bit-field.}
\end{figure}

Since \texttt{FD} dominates and can generate \texttt{SDC}, this subsection examines \texttt{SDC} severity in MatMul across floating-point precisions using the average number of corrupted outputs and RMSE, to understand how precision affects fault propagation through the micro-architecture and the resulting output corruption. The results are shown in \cref{fig:FI_results_SDC}.

% \paragraph{Observation 4:} Transient SEUs on vector operands don't trigger any System Crash, and the resulting SDCs can go unnoticed, making their impact on output correctness particularly severe.
% \paragraph{Observation 5:} FP8 exhibits at least an order-of-magnitude lower SDC incidence than FP16, FP32 and BP16 in our experiments. This indicates that reduced precision might offer further fault-tolerance advantages beyond its reduced encoding budget.

% \paragraph{Observation 6:} The single-value deviation magnitude measured by RMSE follows the same trend as the maximum representable numeric range of each data format.

% \paragraph{Observation 7:} From \cref{fig:FI_results_SDC}, the most severe \texttt{SDC} deviations are concentrated in exponent-targeted faults, with two particularly dominant outliers corresponding to FP32 and BP16. The associated RMSE values reach \textbf{3.40E+37} and \textbf{9.86E+36}, respectively, far exceeding the deviations observed for the other bit-fields and precisions.

% \noindent\textbf{Design hint 2:} These results indicate that the exponent paths of FP32 and BP16 are the highest-leverage protection targets in the design. Prioritizing selective hardening for these cases offers an especially favorable cost-benefit tradeoff, as it directly mitigates the largest observed \texttt{SDC} deviations with limited additional area and power cost.

No \texttt{FS} was triggered during these experiments; instead, SEUs affecting the FPU operand path inside the VFU of the vector co-processor manifested as \texttt{SDC}. As shown in \cref{fig:FI_results_SDC}, the observed RMSE trend follows the numerical range of the format: FP32 is the least resilient, while FP16 shows substantially lower error severity than FP32 and BP16 due to its smaller range. Widening MatMul further improves FP16 robustness, reducing both the number of corrupted outputs and the corresponding RMSE, because the accumulation is carried out in FP32 rather than FP16. This indicates that a wider accumulator can mitigate the propagation of numerical errors induced by injected faults.
FP8 exhibits the lowest output impact, both in terms of corruption spread and RMSE, confirming that lower-precision formats provide an inherent resilience advantage. However, widening is less effective for FP8 than for FP16, since accumulation is performed in 16-bit rather than 32-bit precision.

The most severe \texttt{SDC} events are induced by exponent-targeted faults, with the two dominant worst-case outliers in \cref{fig:FI_results_SDC} corresponding to FP32 and BP16. This suggests that protection should be concentrated on the highest-impact cases rather than applied uniformly across all precisions and bit-fields. In particular, selectively protecting exponent-targeted faults in FP32 and BP16 is especially attractive, as it directly targets the largest observed \texttt{SDC} deviations while requiring protection for only a small subset of cases. By contrast, mantissa faults are often much more tolerable, especially at low precision such as FP8.

% Conclusion and Future Work
\section{Conclusion and Future Work}
We presented an RTL fault-injection sensitivity study of the open-source Spatz vector cluster under SET and SEU fault models. Across 100,000 injections on MatMul and Widening MatMul, faulty data dominates among manifesting outcomes (86\% for SET and 91\% for SEU), with SET hotspots concentrating in the vector execution subsystem and SEU hotspots dominated by the TCDM. We further quantified SDC incidence and numerical severity across FP32/FP16/FP8/BP16, observing at least an order-of-magnitude fewer SDC events in FP8 and substantially higher SDC severity when exponent bits are corrupted. These findings motivate selective enhancement: protect the highest-leverage datapaths (VRF/VLSU interfaces) and apply exponent-focused checking to reduce worst-case deviation at low overhead. Future work will extend the same methodology to additional kernels and broaden fault-site coverage.

\bibliographystyle{ACM-Reference-Format}
\bibliography{refs}


\end{document}